\newcommand{\nc}{\newcommand}
\nc{\kms}{\,{km\,s$^{-1}$}}
\nc{\sgra}{Sgr A}
\nc{\sgrastar}{Sgr $\rm {A}^{*}$}
\nc{\sgraeast}{Sgr A East}
\nc{\sgrawest}{Sgr A West}
\nc{\sgracomp}{Sgr A Complex}
\nc{\as}{\arcsec}
\nc{\HI}{H\,{\sc i}}
\nc{\HII}{H\,{\sc ii}}
\nc{\CI}{C\,{\sc i}}
\nc{\hto}{H$_{2}$O}
\nc{\htmo}{H$_{2}^{16}$O}
\nc{\htio}{H$_{2}^{18}$O}
\nc{\am}{\arcmin}
\nc{\ciso}{C$^{18}$O}
\nc{\hctn}{HC$_{3}$N}
\nc{\htwo}{H$_{2}$}
\nc{\ot}{O$_{2}$}
\nc{\nht}{NH$_{3}$}
\nc{\htco}{H$_{2}$CO}
\nc{\htre}{H$_{3}$O${^+}$}
\nc{\ohs}{OH-Streamer}
\nc{\msol}{{$\mathrm{M}_{\odot} $}}
\nc{\met}{CH$_{3}$OH}
\nc{\metiso}{$^{13}$CH$_{3}$OH}
\nc{\fas}{$\farcs$}
\nc{\chl}{H$_{2}$Cl$^{+}$}

\documentclass[bibyear]{aa}

\usepackage{graphicx}
\usepackage{txfonts}

\begin{document}

\title{{\it Herschel} HIFI observations of the \sgra\ +50 \kms\ Cloud 
\thanks{{\it Herschel} is an ESA space observatory with science instruments
  provided by European-led Principal Investigator consortia and with
  important participation from NASA.}}

\subtitle{Deep searches for \ot\ in emission and foreground absorption}

\author{Aa.\,Sandqvist\inst{1}
  \and B.\,Larsson\inst{1}
  \and \AA .\,Hjalmarson\inst{2} 
\and P.\,Encrenaz\inst{3}
\and M.\,Gerin\inst{4} 
\and P.\,F.\,Goldsmith\inst{5}
\and D.\,C.\,Lis\inst{3,6}
\and R.\,Liseau\inst{2}
\and L.\,Pagani\inst{3}
\and E.\,Roueff\inst{7}
\and S.\,Viti\inst{8}
}

\institute{Stockholm Observatory, Stockholm University, AlbaNova
  University Center, SE-106 91 Stockholm, Sweden\\ \email{aage@astro.su.se}
\and Department of Earth and Space Sciences, Chalmers University of
Technology, Onsala Space Observatory, SE-429 92 Onsala, Sweden
\and LERMA, Observatoire de Paris, PSL Research University, CNRS,
Sorbonne Universités, UPMC Univ. Paris 06, F-75014, Paris, France 
\and LERMA, Observatoire de Paris, PSL Research University, CNRS,
Sorbonne Universités, UPMC Univ. Paris 06, École normale supérieure,
F-75005, Paris, France 
\and Jet Propulsion Laboratory, California Institute of Technology,
4800 Oak Grove Drive, Pasadena CA 91109, USA
\and California Institute of Technology, Cahill Center for Astronomy
and Astrophysics 301-17, Pasadena, CA 91125, USA
\and LERMA, Observatoire de Paris, PSL Research University, CNRS,
Sorbonne Universités, UPMC Univ. Paris 06, F-92190, Meudon, France  
\and Department of Physics and Astronomy, University College London,
London WC1E 6BT, UK 
}

\offprints{\\ Aage Sandqvist, \email{aage@astro.su.se}}

\date{Received $<$date$>$; accepted $<$date$>$}

\abstract {The {\it Herschel} Oxygen Project (HOP) is an Open Time
  Key Program, awarded 140 hours of observing time to search for
  molecular oxygen (\ot) in a number of interstellar sources. To date
  \ot\ has definitely been detected in only two sources, namely $\rho$ 
  Oph A and Orion, reflecting the extremely low abundance of
  \ot\ in the interstellar medium.} {One of the sources in the HOP
  program is the +50 \kms\ Cloud in the \sgracomp\ in the centre of the
  Milky Way. Its environment is unique in the Galaxy and this property
is investigated to see if it is conducive to the presence of \ot.}
{The {\it Herschel} Heterodyne Instrument for the Far Infrared (HIFI)
  is used to search for the 487 and 774 GHz emission lines of \ot.}
  {No \ot\ emission is detected towards the \sgra\ +50 \kms\ Cloud, but a
  number of strong emission lines of methanol (\met) and absorption
  lines of chloronium (\chl) are observed.}  
  {A $3\sigma$ upper limit for the fractional abundance ratio of
    [\ot]/[\htwo] in the \sgra\ +50 \kms\ Cloud is found to be $X$(\ot) $\le 5
    \times 10^{-8}$. However, since we can find no other realistic molecular
    candidate than \ot\ itself, we very tentatively suggest that two
    weak absorption lines at 487.261 and 487.302 GHz may be caused by
    the 487 GHz line of \ot\ in two foreground spiral arm
    clouds. By considering that the absorption may only be apparent, the
    estimated upper limit to the \ot\ abundance of  
    $\le (10 - 20) \times 10^{-6}$ in these foreground clouds is very
    high, as opposed to the upper limit in the \sgra\ +50 \kms\ Cloud
    itself, but similar to what has been reached in recent chemical
    shock models for Orion. This abundance limit was determined also
    using {\it Odin} non-detection limits, and assumes that
      \ot\ fills the beam. If the absorption is due to a
      differential {\it Herschel} OFF-ON emission, the \ot\ fractional
      abundance may be of the order of $\approx (5 - 10) \times
      10^{-6}$. With the assumption of pure absorption by
        foreground clouds, the unreasonably high abundance of $(1.4 -
        2.8) \times 10^{-4}$ was obtained. The rotation temperatures
        for \met-$A$ and \met-$E$ lines in the +50 \kms\ Cloud are
        found to be $\approx 64$ and 79 K, respectively, and the
        fractional abundance of \met\ is approximately $5 \times
        10^{-7}$.}

\keywords{Galaxy: centre -- ISM: individual objects: \sgra\  -- ISM:
  molecules -- ISM: clouds}

\titlerunning{{\it Herschel} observations of the \sgra\ +50 \kms\ Cloud}
\authorrunning{Aa. Sandqvist et al.}
\maketitle

\section{Introduction}

The core of the Milky Way Galaxy is a region of great complexity
containing a wide variety of physical environments. At the very
centre resides a four-million-solar-mass supermassive black hole,
whose non-thermal radio continuum signature is called
\sgrastar. Orbiting around it, at a distance of one to a few pc, with
a velocity of about 100 km/s, is a molecular
torus called the Circum Nuclear Disk (CND). The CND has a mass of
$10^4$ to $10^5$ \msol\ and a gas temperature of several hundred
degrees. Beyond this, there exists a large molecular belt, consisting
predominantly of two giant molecular clouds (GMC)s, called the +50 and
the +20 \kms\ Clouds. Both GMCs are massive, about $5 \times 10^5$
\msol, with a density of 
$10^4$ - $10^5$ cm$^{-3}$, a gas temperature of 80 - 100 K, and a dust
temperature of 20 - 30 K (e.g. Sandqvist et al. \cite{san08}). The +50
\kms\ Cloud, as depicted in the CS molecule, is shown in Fig. 1. A
giant energetic ($ > 10^{52}$ erg) supernova remnant (SNR)-like
nonthermal continuum radio 
shell (diameter about 8 pc), known as \sgraeast, is plowing into this
molecular belt from the side near \sgrastar, creating regions of shock
interaction, especially at the inner western surface of the +50 \kms\
Cloud. Near the opposite eastern side of the +50 \kms\ Cloud 
four compact \HII\ regions exist, named A - D by Ekers et
al. (\cite{eke83}). General reviews of the Galactic centre have been
presented by e.g. Mezger et al. (\cite{mez96}) and Morris \& Serabyn
(\cite{mor96}), with an up-to-date introduction to the \sgracomp\ given
by Ferri\`ere (\cite{fer12}).

\begin{figure}
  \resizebox{\hsize}{!}{\rotatebox{90}{\includegraphics{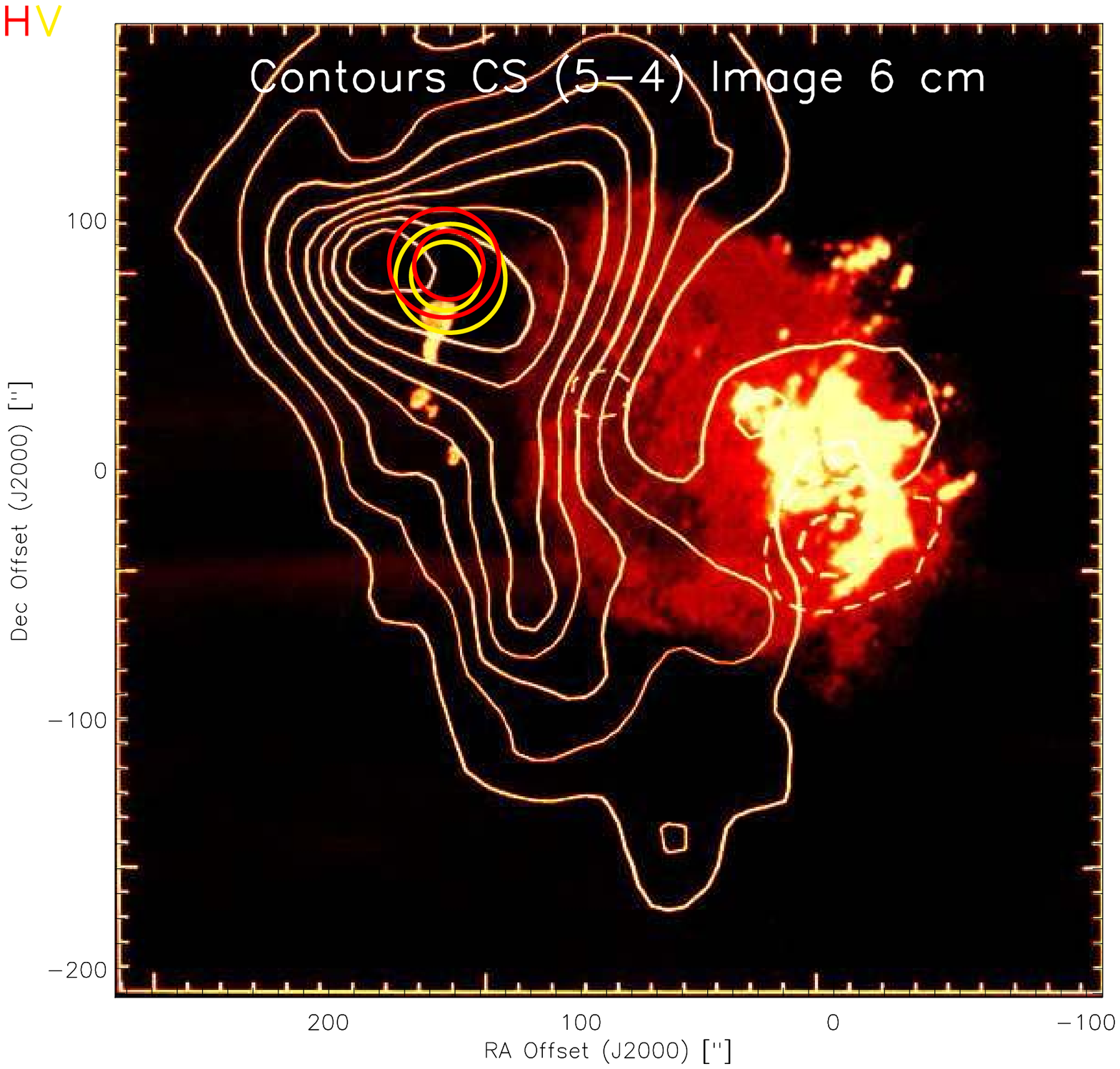}}}
  \caption{\sgra\ +50 \kms\ Cloud, delineated by the integrated
emission (+20 to +80\kms) in the CS(5-4) line (contours),
superimposed upon a 6 cm continuum map of the Sgr A Complex (colour
image - red: nonthermal \sgraeast, yellow: thermal \sgrawest\ and the four
compact \HII\ regions of which A is the northernmost, close to our
observational position) (Serabyn et al. \cite{ser92}). The FWHM {\it
  Herschel} 487 and 774 GHz beams, H (northernmost) and V
polarizations, for the HOP observations we have carried out are indicated by
red and yellow circles, respectively.}  
  \label{1}
\end{figure}

The line of sight from the Sun to \sgra\ crosses a number of Galactic
spiral arms and thus can serve to probe the properties of the
Local Arm-Feature A (+5 \kms), the Sagittarius Arm ($-2$ \kms), the
giant Scutum Arm ($-7$ \kms), the $-30$ \kms\ Arm ($-30$ \kms), the 3
kpc Arm ($-52$ \kms), and the Expanding Molecular Ring (EMR) ($-130$
\kms), with the approximate radial velocities with respect to the
local standard of rest (LSR) given in parenthesis (e.g. Sandqvist
\cite{san70}). This configuration is sketched in Fig. 2. The physical
conditions in the spiral arms are radically different from those in
the Galactic centre region. The arms contain a diffuse cold neutral
medium with densities of the order of $50 - 100$ cm$^{-3}$ and a
kinetic temperature of about 100 K, in approximate pressure equilibrium
with a warm neutral medium with densities of about 0.4 cm$^{-3}$ and a
temperature of about 8000 K (Gerin et al. \cite{ger15}). The arms
  also contain cold dense matter, often in the form of dark clouds,
  with densities of the order of $10^3 - 10^4$ cm$^{-3}$ and a kinetic
  temperature of about 10 K. Absorption lines from regions seen
towards background continuum sources serve as convenient probes of the
properties of the cold neutral medium in the spiral arms,
(e.g. Karlsson et al. \cite{kar13}).  

\begin{figure}
  \resizebox{\hsize}{!}{\rotatebox{0}{\includegraphics{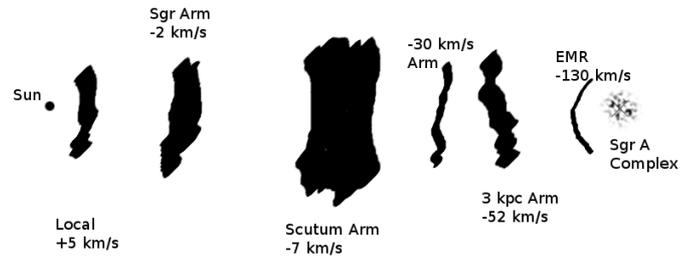}}}
  \caption{Line of sight from the Sun to the Sgr A complex at
    the Galactic centre cuts through a number of spiral arm features,
    shown in this sketch, which also indicates their approximate radial
    velocities with respect to the LSR. Assuming that the distance
    from the Sun to the Galactic centre is 8 kpc, the estimated distances from
    the Sun to the spiral arm features are as follows: Local - $\frac{1}{2}$
    kpc, Sgr Arm - $1\frac{1}{2}$ kpc, Scutum Arm - 3 kpc, -30 km/s Arm - 4 kpc,
    3 kpc Arm - 5 kpc, EMR - $7\frac{1}{2}$ kpc.}  
  \label{2}
\end{figure}

The search for molecular oxygen (\ot) in the interstellar medium
was a major objective for the Submillimeter Wave Astronomy
Satellite (SWAS) (Melnick et al. \cite{mel00}) and the 
{\it Odin} satellite (Nordh et al. \cite{nor03}). The surprising
results were that the abundance of molecular oxygen was more than 100
times lower than expected from gas-phase chemistry predictions, and
only in one source, namely $\rho$ Oph A, was \ot\ actually detected
(Larsson et al. \cite{lar07}). {\it Odin} also led to an \ot\ 
abundance upper limit of $\lesssim 1.2 \times 10^{-7}$ for the whole \sgra\
molecular belt region and the results of a first simple modeling are
presented by Sandqvist et al. (\cite{san08}) and Hollenbach et
al. (\cite{hol09}). (The {\it Odin} 119-GHz \ot\ 9 arcmin beam included both the
+20 and +50 \kms\ Clouds and the ridge in between).     

The {\it Herschel} Open Time Key Program, {\it Herschel} Oxygen
Project (HOP), (co-principal investigators: P. Goldsmith and R. Liseau)
was designed to 
investigate the problem of the unexpectedly low \ot\ abundance and we
have observed a number of sources with {\it Herschel}. The first results,
reporting the detection of \ot\ in Orion, were published by
Goldsmith et al. (\cite{gol11}), who also give a thorough review of
the question of the abundance of \ot. The second HOP results,
confirming the {\it Odin} detection of \ot\ in $\rho$ Oph A, were
presented by Liseau et al. (\cite{lis12}). Observations towards the
  Orion Bar photo-dissociation region (PDR) and the low-mass protostar
  NGC 1333 IRAS 4A did not 
  reveal the presence of any \ot\ (Melnick et al. \cite{mel12}; Yildiz
  et al. \cite{yil13}). Recently, Chen et
al. (\cite{che14}) developed a shock model to explain the \ot\
observations of the Orion \htwo\ Peak 1 and Hot Core. One of the additional
objects observed by HOP is the \sgra\ +50 \kms\ Cloud, the topic of
the present paper. In addition to improving the understanding
of the \sgracomp, we chose \sgra\ rather than Sgr B2 - the more
standard molecular search candidate - partly beacuse of the greater variety of
physical phenomena in the \sgracomp, which  possibly increases the
chances of success in searching for this elusive molecule. In addition,
the expected lower line density towards \sgra\ should facilitate
more secure identification of the molecular line carriers.

\section{Observations}

The {\it Herschel} \ot\ search in the \sgracomp\ was focused on the
+50 \kms\ Cloud. The chosen position, J2000.0 $17^h45^m51\fs70,
-28\degr59\arcmin09\farcs0$ (+152\as,+75\as\ - equatorial offsets 
from \sgrastar ), corresponds to the position used in the {\it Odin}
\hto/\ot\ observations (see Table 1 and Fig. 1 of Sandqvist et 
al. \cite{san08}) and is indicated with the {\it Herschel}
beams in Fig. 1. This position lies in
the region not yet reached by the shock caused by \sgraeast. Thus it
is reasonable that the gas there has a lower temperature,
which is also implied by the lack of significant \nht(3,3) emission
(McGary et al. \cite{mcg01}). This position is also very close to the
northernmost of the four compact \HII\ regions (A) - see Fig. 1.  

The 487 and 774 GHz observations were performed on OD 503 (2010
September 29) and OD 491 (2010 September 16-17), respectively. The
double beam-switch mode (DBS) was used yielding two
OFF-beam positions located 3\am\ on either side of the source
along a position angle (PA) of 92\degr. The OFF-positions thus fall
outside the main molecular emission region. To ensure the reality of features
without confusion from conflicting upper and lower sideband effects, a
sequence of eight local oscillator (LO) settings was used with
separations from 120 to 250 MHz and from 30 to 270 MHz for the 487 and
774 GHz observations, 
respectively. This resulted in a total ON-source integration time of 2.9
hours for each of the two frequencies. After combining the H- and
V-polarisation channels, the final value of the ON-source integration
time is 5.8 hours. The Wide Band Spectrometer (WBS) has a channel
resolution of 1.10 MHz, or 0.68 and 0.43 \kms at 487 and 774 GHz,
respectively. The standard {\it Herschel} pipeline,
HIPE\footnote{http://www.cosmos.esa.int/web/herschel/hipe-download}
(Ott \cite{ott10}) version 11, has been employed for calibration and
initial reductions.  

Throughout the paper the intensities are given in terms of the
antenna temperature, $T_{\rm A}$, unless otherwise stated. The main
beam efficiencies for the {\it Herschel} telescope are 0.62 in the
487, 498 and 0.63 in the 774, 785 GHz frequency bands, respectively,
while the half-power beamwidths are 43\as\ and 27\as\ (Mueller et
al. \cite{mue14}; Roelfsema et al. \cite{roe12}). 

\section{Results}

We present here our HIFI observations towards the +50 \kms\ Cloud in
\sgra. The 487 and 498 GHz spectra are shown in Figs. 3 and 4,
respectively, and the 774 and 785 GHz spectra in Figs. 5 and 6. They
are dominated by a number of \met\ lines,  
which we show in more detail in Fig. 7, together with the higher
frequency \met\ lines. The two double absorption features seen in
Fig. 3, which we have now identified as originating from H$_{2}^{35}$Cl$^{+}$ and
H$_{2}^{37}$Cl$^{+}$, have previously been presented by Neufeld et
al. (\cite{neu12}). There are many other emission lines and an
additional absorption line which we have identified in our four
spectra. The absorption line, U1, seen in the 774 GHz band, has a
shape remarkably similar to the 0\kms\ feature of the \chl\ line, see
Fig. 8. Assuming that the U1-line arises from the same region as the
\chl\ line, we determine a frequency to be 771.365 GHz,
which is the rest frequency of the $5_{4,1} - 6_{3,4}$
transition of the deuterated formaldehyde molecule, HDCO. For
comparison, we also display in Fig. 8 the relevant part of the 6 cm 
formaldehyde (\htco) absorption profile obtained towards \sgra\ with
the National Radio Astronomy Observatory (NRAO) 140-foot radio
telescope (Sandqvist \cite{san70}). However, 
the high lower state energy value of 137 K, together with the fact
that neither the corresponding high energy absorption line from \htco\ itself
nor that of H$_2^{13}$CO have been detected toward the Sgr A +50 \kms\
Cloud, make the identification of such an absorption line from HDCO very
tentative.

A summary of the emission lines and Gaussian fits is presented in
Table 1, where $V$, $T_{\rm A}$ and $\Delta V$ are the fitted radial
velocity with respect to the LSR, antenna temperature, and full width
at half measure (FWHM)
line width, respectively. Also, the integrated line intensity,
$\int{T_{\rm A}{\rm d}V}$, is listed. In a few cases, the lines are
blended and the integrated intensity is therefore determined from the
Gaussian parameters. The Einstein coefficient $A_{ul}$ and energy of
the upper level $E_{u}$ for most of the  transitions are also
listed. Since methanol is a slightly asymmetric rotor, the two $A-$ and
$E-$ symmetric states have been identified. A summary of the absorption
lines is presented in Table 2. The line identifications rely upon the
Splatalogue\footnote{http://www.cv.nrao.edu/php/splat/}, the Cologne
Database for Molecular Spectroscopy
(CDMS\footnote{http://www.astro.uni-koeln.de/cdms}) (M\"uller et
al. \cite{mul05}), and the Jet Propulsion Laboratory
(JPL\footnote{http://spec.jpl.nasa.gov/}) (Pickett et
al. \cite{pic98}) molecular spectroscopy data bases available online.    

\begin{figure*}
  \resizebox{\hsize}{!}{\rotatebox{270}{\includegraphics{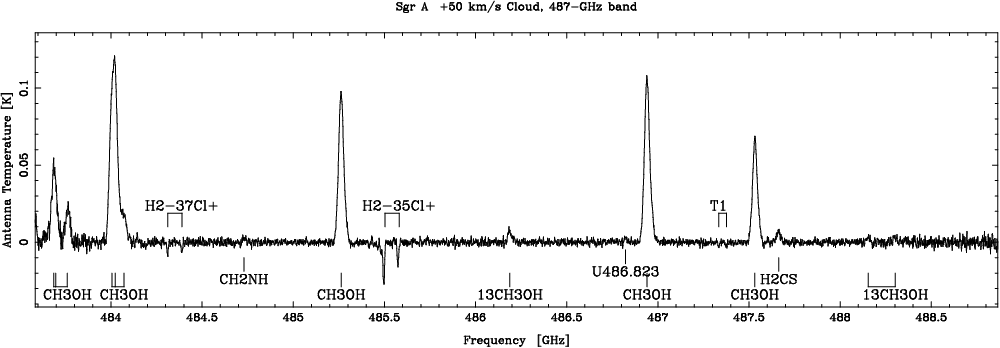}}}
  \caption{487 GHz spectrum observed towards the \sgra\ +50 \kms\
    Cloud, with a linear baseline subtracted. The velocity resolution
    is 0.9 \kms. The frequency scale is calibrated for a radial
    velocity of 45.2 \kms\ for easier identification of the emission
    lines. (This results in a shift of the frequencies of  the
    absorption lines, which do not originate in the +50 \kms\ Cloud,
    see text).}   
  \label{3}
\end{figure*}

\begin{figure*}
  \resizebox{\hsize}{!}{\rotatebox{270}{\includegraphics{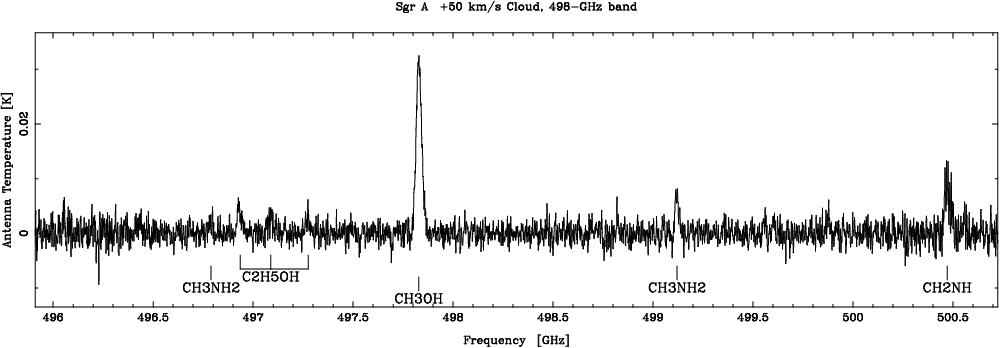}}}
  \caption{498 GHz spectrum observed towards the \sgra\ +50 \kms\ Cloud,
    with a linear baseline subtracted. The velocity resolution is 0.9
    \kms. The frequency scale is calibrated for a radial velocity of
    45.2 \kms\ for easier identification of the emission lines.}  
  \label{4}
\end{figure*}

\begin{figure*}
  \resizebox{\hsize}{!}{\rotatebox{270}{\includegraphics{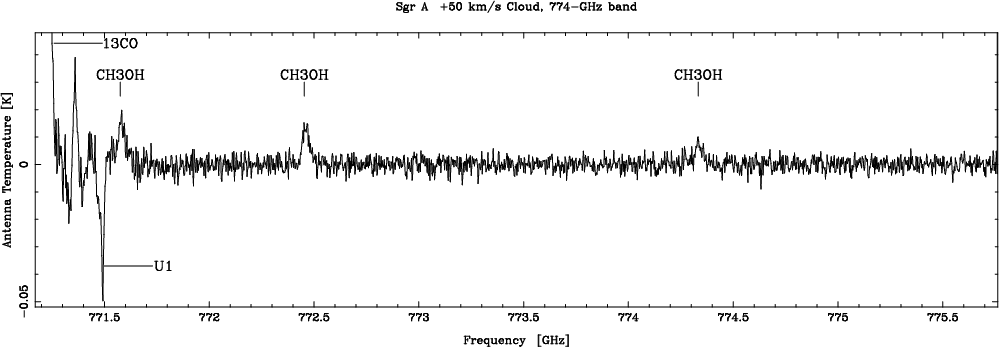}}}
  \caption{774 GHz spectrum observed towards the \sgra\ +50 \kms\
    Cloud, with a linear baseline subtracted. The velocity resolution
    is 1.0 \kms.The frequency scale is calibrated for a radial velocity of
    45.2 \kms\ for easier identification of the emission lines. (This
    results in a shift of the frequency of the U1 absorption line,
    which does not originate in the +50 \kms\ cloud, see text). }    
  \label{5}
\end{figure*}

\begin{figure*}
  \resizebox{\hsize}{!}{\rotatebox{270}{\includegraphics{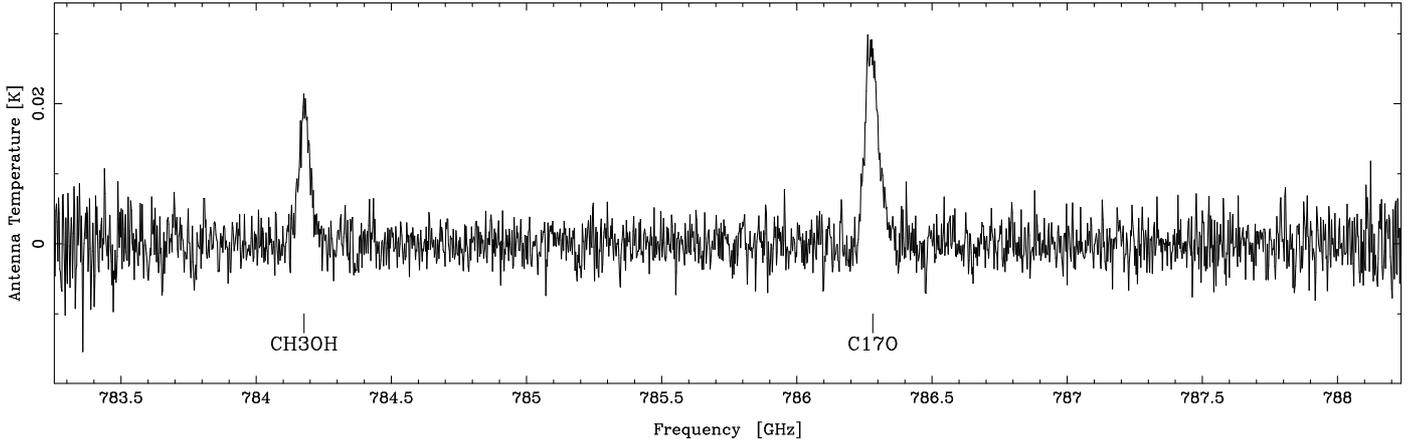}}}
  \caption{785 GHz spectrum observed towards the \sgra\ +50 \kms\
    Cloud, with a linear baseline subtracted. The velocity resolution
    is 1.0 \kms. The frequency scale is calibrated for a radial velocity of
    45.2 \kms\ for easier identification of the emission lines.}   
  \label{6}
\end{figure*}

\begin{figure}
  \resizebox{\hsize}{!}{\rotatebox{0}{\includegraphics{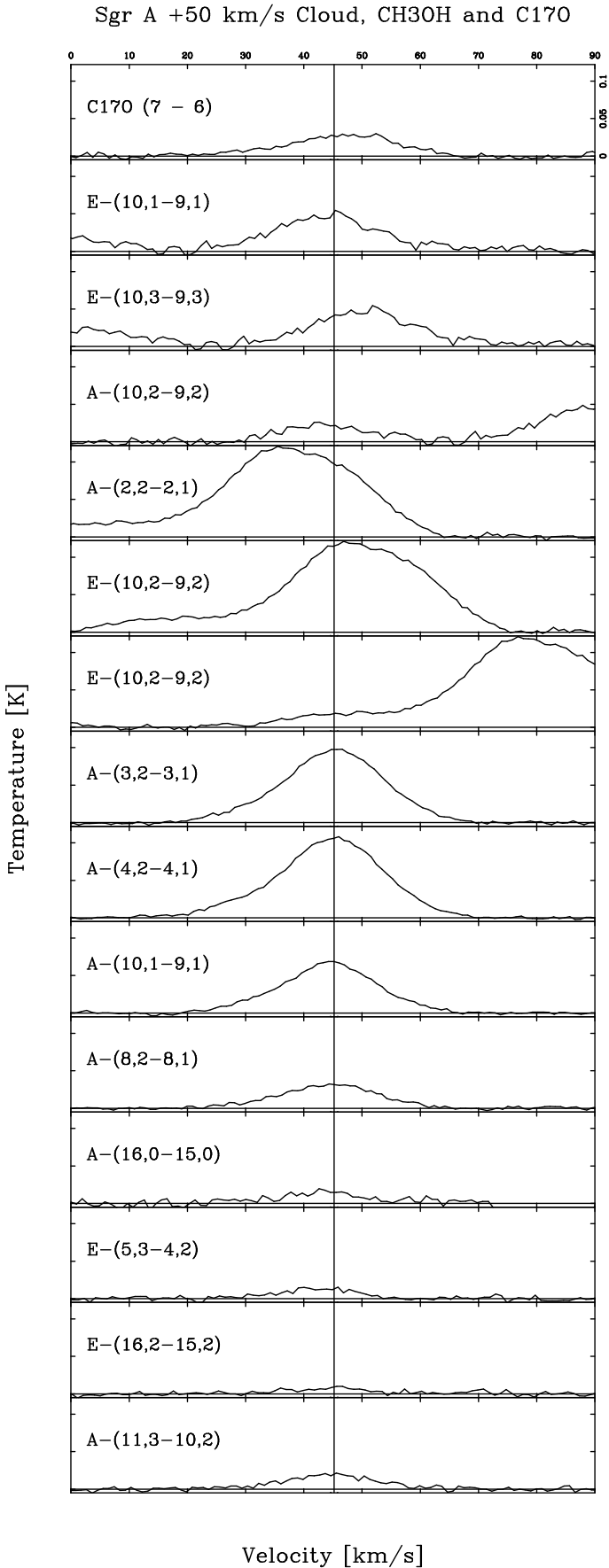}}}
  \caption{All the \met\ emission lines in the four spectral bands and
    the C$^{17}$O ($7-6$) emission line towards the \sgra\ +50 \kms\
    Cloud.} 
  \label{7}
\end{figure}

\begin{figure}
  \resizebox{\hsize}{!}{\rotatebox{270}{\includegraphics{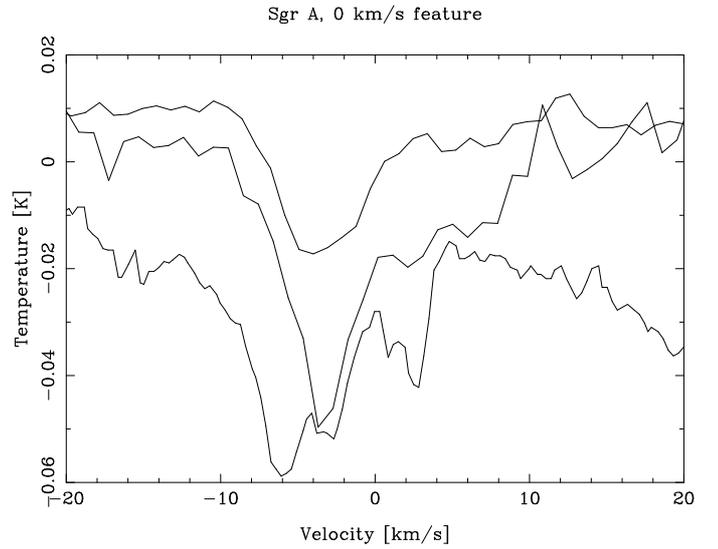}}}
  \caption{{\it Upper profile}: {\it Herschel} H$_2^{35}$Cl$^+$
    absorption profile, raised by 0.01 K for clarity. {\it Middle
      profile}: {\it Herschel} Unidentified, or HDCO?, absorption line
    profile. {\it Lower profile}: 6 cm \htco\ absorption profile
    (6 arcmin resolution, Sandqvist 1970); this profile has been
    scaled by a factor of 0.02.}     
  \label{8}
\end{figure}

\begin{figure}
  \resizebox{\hsize}{!}{\rotatebox{0}{\includegraphics{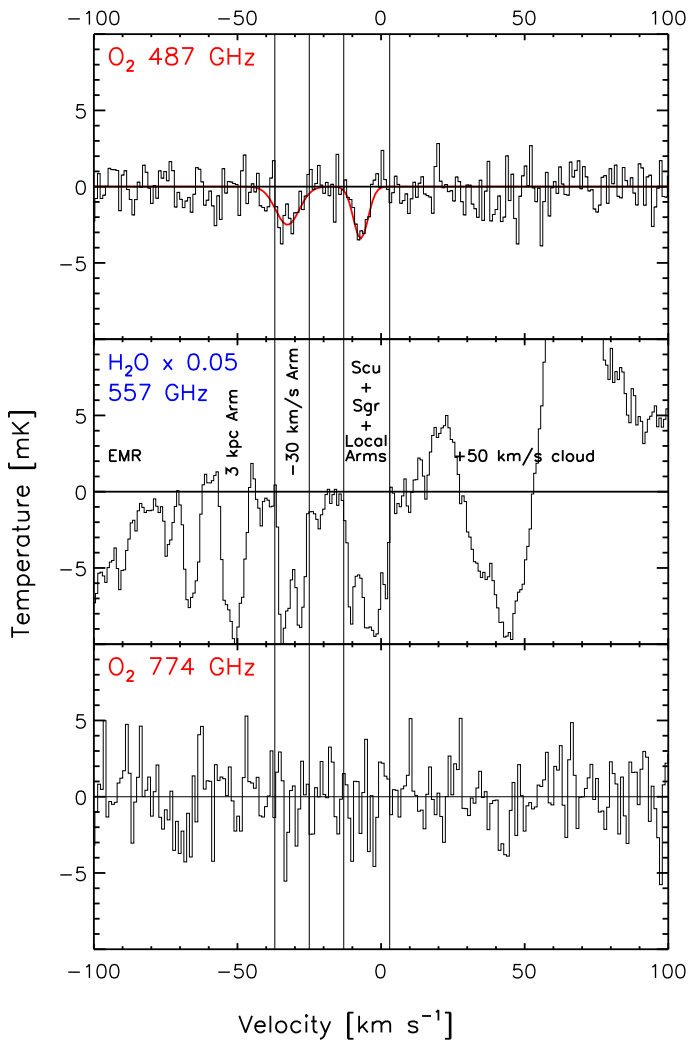}}}
  \caption{487 and 774 GHz \ot\ spectra towards the \sgra\ +50
    \kms\ Cloud, compared with a 557 GHz o-\hto\ profile observed
    44\as\ west and 20\as\ south of our position. The scale of the \hto\
    profile has been multiplied by 0.05. The channel resolution of
    both \ot\ spectra is 0.9 \kms.}   
  \label{9}
\end{figure}

\begin{table*}
\caption{Summary of observed emission lines and Gaussian fits.}
\begin{flushleft}
\begin{tabular}{lllcrllll}
\hline\hline\noalign{\smallskip}
Species & Transition & Frequency & $A_{ul}$ & $E_{u}$ & $V$  & $T_{\rm
  A}$  & $\Delta V$  & $\int{T_{\rm A}{\rm d}V}$   \\
    &     &   (GHz)   & (s$^{-1}$) & (K) &   (\kms)   &   (K)   &
    (\kms)  & (K \kms) \\ 
\hline\noalign{\smallskip}
\met-$E$ & $10_{1,9} - 9_{1,8}$  &  483.686 & $5.13 \times 10^{-4}$ & 148.7 &
46.1 &  0.0312 & 18.4 & 0.574 \tablefootmark{a} \\
\met-$E$ & $10_{3,7} - 9_{3,6}$  &  483.697 & $4.62 \times 10^{-4}$ &
175.4 & 45.0 &  0.0198 & 19.1 & 0.378 \tablefootmark{a} \\
\met-$A$ & $10_{2,8} - 9_{2,7}$  &  483.761 & $4.90 \times 10^{-4}$
& 165.4  & 43.8 &  0.0227 & 18.5 &  0.428 \\
\met-$A$ & $2_{2,1} - 2_{1,2}$   &  484.005 & $4.01 \times 10^{-4}$ &
44.7 & 46.4 &  0.0702 & 18.1 &
1.271 \tablefootmark{a} \\
\met-$E$ & $10_{2,8} - 9_{2,7}$  &  484.023 & $4.83 \times 10^{-4}$ &
150.0 & 44.6 &  0.0982 & 17.6 & 1.728 \tablefootmark{a} \\
\met-$E$ & $10_{2,9} - 9_{2,8}$  &  484.072 & $4.89 \times 10^{-4}$
& 153.6 & 46.5 &  0.0208 & 20.8 & 0.433 \tablefootmark{a} \\
CH$_2$NH & $2_{2,0} - 2_{1,1}$ &  484.729 & $4.85 \times 10^{-4}$ &
40.7 & 43.6 & 0.0029 & 11.2 &  0.039 \\  
\met-$A$ & $3_{2,2} - 3_{1,3}$   &  485.263 & $5.05 \times 10^{-4}$ &
51.6 & 45.2 &  0.0955 & 20.6 &  2.098 \\ 
\metiso & $7_{0,7} - 6_{1,6}$   &  486.188 & $3.02 \times
10^{-4}$ & 76,5 & 45.4 &  0.0070 & 16.1 & 0.122 \tablefootmark{b} \\ 
U486.823   &       &  486.823  &    &    &  44.8  &  0.0034  &  8.6  &
0.033 \\ 
\met-$A$ & $4_{2,3} - 4_{1,4}$   &  486.941 & $5.51 \times 10^{-4}$ &
60.9 & 44.8 &  0.1038 & 21.3 & 2.386  \\
\met-$A$ & $10_{1,9} - 9_{1,8}$  &  487.532 & $5.15 \times 10^{-4}$
& 143.3 & 44.3 &  0.0654 & 19.5 & 1.363  \\
H$_2$CS & $14_{1,13} - 13_{1,12}$  &  487.663 & $1.77 \times 10^{-3}$
& 188.8 & 46.1 & 0.0068 &  16.3 & 0.117  \\  
\metiso & $4_{1,4} - 3_{0,3}$ & 488.154 & $7.46 \times 10^{-4}$&
37.0 & 44.7 & 0.0033 & 11.7 & 0.037  \\ 
\metiso & $4_{2,3} - 4_{1,4}$ & 488.302 & $5.50 \times 10^{-4}$ &
60.4 & 45.4 & 0.0035 &  8.3 & 0.040  \\ 
CH$_3$NH$_2$ & $9_{1,6} - 8_{0,7}$ & 496.790 & & 100.9 & 47.9 & 0.0018
& 14.8 & 0.023  \\ 
C$_2$H$_5$OH & $15_{5,10} - 14_{4,11}$ & 496.935 & & 132.2 & 47.0 & 0.0052 &
12.6 & 0.096  \\ 
C$_2$H$_5$OH & $9_{7} - 8_{6}$ & 497.088 &  & 99.2 & 44.5 & 0.0026 & 
18.5 & 0.154  \\ 
C$_2$H$_5$OH & $12_{6,7} - 11_{5,6}$ & 497.276 & & 110.7 & 47.3 &
0.0026 & 15.0 &  0.037 \\ 
\met-$A$       &  $8_{2,7} - 8_{1,8}$ & 497.828 & $6.36 \times
10^{-4}$ & 121.3 & 44.4 & 0.0313 &
19.6 & 0.689  \\ 
CH$_3$NH$_2$ & $9_{1,3} - 8_{0,2}$ & 499.120 & & 100.7 & 45.3 & 0.0072 &
12.5 & 0.099 \tablefootmark{c} \\ 
CH$_3$NH$_2$ & $6_{2,7} - 5_{1,6}$ & 499.868 & &  60.6 & 40.0 & 0.0033 &
10.6 & 0.035 \\ 
CH$_2$NH   & $2_{2,1} - 2_{1,2}$  & 500.471  & $5.09 \times 10^{-4}$ & 40.7 & 44.4 & 0.0086
& 20.8 & 0.187  \\  
$^{13}$CO  & $7 - 6$ & 771.184 &   & 148.0 & & >1 &   &   \\
\met-$A$      & $16_{0,16} - 15_{0,15}$ & 771.576 & $2.10 \times 10^{-3}$
& 315.2 & 43.3 & 0.0138 & 20.1 & 0.310  \\
\met-$E$      & $5_{3,2} - 4_{2,2}$   & 772.454 & $2.70 \times
10^{-3}$ & 82.5 & 42.3 & 0.0136 & 15.2 & 0.217  \\ 
\met-$E$      & $16_{2,14} - 15_{2,13}$ & 774.333 & $2.06 \times
10^{-3}$ & 338.1 &  44.3 & 0.0074 & 17.4 & 0.142  \\
\met-$A$      & $11_{3,8} - 10_{2,9}$ & 784.177 & $1.86 \times
10^{-3}$ & 203.0 & 44.6 & 0.0190 & 18.3 & 0.381  \\ 
C$^{17}$O  & $7 - 6$       & 786.281 & & 151.0 & 46.8 & 0.0281 & 21.7 & 0.653  \\
\noalign{\smallskip}\hline\end{tabular}
\end{flushleft}
 $^a$ From Gaussian fits of overlapping lines (line
   blends) \\
$^b$ Blend with H$_2$C$^{18}$O $(7_{3,4}- 6_{3,3})$ at
  487.187 GHz, $E_{\rm u}= 200.0$ K \\
$^c$ Blend with HDCO $(6_{1,6}- 5_{0,5})$ at
  499.142 GHz, $E_{\rm u}= 70.1$ K \\
\end{table*}

\begin{table*}
\caption{Summary of observed absorption lines and Gaussian fits.}
\begin{flushleft}
\begin{tabular}{lllllllllll}
\hline\hline\noalign{\smallskip}
Species & Transition & Frequency & $V$  & $T_{\rm A}$  & $\Delta V$  &
$\int{T_{\rm A}{\rm d}V}$    \\
    &     &   (GHz)   &   (\kms)   &   (K)   &   (\kms)  & (K \kms) \\ 
\hline\noalign{\smallskip}
H$_2^{37}$Cl$^+$ & $1_{11} - 0_{00}$ &  484.232 & $-51.4$ & $-0.0067$
& 5.7 & $-0.039$   \\
   &    &   & $-3.5$ & $-0.0088$ & 4.9 & $-0.043$ \tablefootmark{a} \\
   &    &   & $+4.2$ & $-0.0006$ & 5.3 & $-0.003$  \tablefootmark{a} \\ 
H$_2^{35}$Cl$^+$ & $1_{11} - 0_{00}$ &  485.418 & $-51.2$ & $-0.0152$ &
5.7 & $-0.091$    \\
   &    &   & $-3.4$ & $-0.0285$ & 6.1 & $-0.174$ \tablefootmark{a} \\
   &    &   & $+5.6$ & $-0.0074$ & 6.7 & $-0.050$ \tablefootmark{a} \\
T1 (\ot ?  & $3_3 - 1_2$) & 487.249 &  $-32.6$ & $-0.0025$
& 7.5 & $-0.018$ \\ 
   &    &   & $-7.1$ & $-0.0034$ & 4.7 & $-0.016$  \\ 
U1 (HDCO? & $5_{4,1} - 6_{3,4}$) & 771.365 & $-3.6$ & $-0.0459$ & 5.3
& $-0.243$ \tablefootmark{a} \\
   &    &   & $+3.8$ & $-0.0201$ & 8.9 & $-0.179$ \tablefootmark{a} \\
\noalign{\smallskip}\hline\end{tabular}
\end{flushleft}
$^a$ From Gaussian fits \\
$^{ }$ T and U stand for Tentative and Unidentified, respectively
\end{table*}

We see no significant signs of any {spectral emission features} at the
O$_2$ frequencies of 
487.249 and 773.840 GHz in Figs. 3 and 5, respectively. We present
these two spectra near the \ot\ frequencies in greater detail in
Fig. 9. To clarify some of the expected spectral features, often seen in
molecular lines towards the \sgra\ +50 \kms\ Cloud, we have also
plotted a relevant part of a 557 GHz o-\hto\ profile, observed with
{\it Herschel} towards a position which is 44\as\ west and 20\as\
south of ours. Other comparisons with OH, \hto, NH$_3$, and C$^{18}$O
are possible, see e.g. Karlsson et al. (\cite{kar13}).

For the +50 \kms\ Cloud, we obtain mean values of the radial
velocity and FWHM line width of 28
emission lines presented in Table 1 
as $V_{\rm mean}=+45.2$ \kms\ and $\Delta V_{\rm mean}=17.1$ \kms. To
obtain an upper limit for the expected \ot\ emission line strengths, we smooth
the spectra to a velocity resolution of 17.1 \kms\ and determine the
root mean square (RMS) value of the resulting spectra. These values
are 0.22 and 0.42 mK 
for the 487 and 774 GHz lines, respectively. Correcting for the beam
efficiencies of 0.62 and 0.63 (Mueller et al. \cite{mue14})  and
obtaing 3$\sigma$ values we get $T_{{\rm mb}, 3\sigma} \le
1.06$ mK for the 487 GHz and $\le 2.0$ mK for the 774 GHz \ot\ lines.

\section{Discussion}
\subsection{The +50 \kms\ Cloud}

We derive an estimate of the upper limit to the abundance of \ot\ in the +50
\kms\ Cloud, using the online version of RADEX
\footnote{http://www.sron.rug.nl/ $\tilde{}$ vdtak/radex/radex.php}
(van der Tak et al. \cite{van07}). Assuming a gas temperature of $T=80$ K and
a gas density of $n_{\rm H_2}=10^4$ cm$^{-3}$ (Walmsley et al. \cite{wal86};
Sandqvist et al. \cite{san08}) and a line width of 17.1 \kms\ (see Sect. 3),
we apply RADEX to our two upper limits of $T_{{\rm mb}, 3\sigma}$  for
the 487 and 774 GHz \ot\ lines. The resulting
upper limit for the \ot\ column density is $N$(\ot) $\le 1.4 \times
10^{16}$ cm$^{-2}$. With a column density of H$_2$
towards the +50 \kms\ Cloud $N$(H$_2$) = $2.4 \times 10^{23}$
cm$^{-2}$ (Lis \& Carlstrom \cite{lis94}), we finally obtain a
3$\sigma$ upper limit for the \ot\ abundance with respect to H$_{2}$
of $X$(\ot) $\le 5 \times 10^{-8}$. These column density and abundance
limits are lower than the {\it Odin} limits for this
  source and also somewhat lower than the level predicted by the
extended PDR model of Hollenbach et al. (\cite{hol09}), as discussed
by Sandqvist et al. (\cite{san08}). The abundance limit is also
lower than {\it Odin} limits obtained for a dozen Galactic
sources of different types (Pagani et al. \cite{pag03}). 

\subsection{\ot\ absorption lines?}

The 487 GHz profile in Fig. 9 indeed shows no sign of any \ot\
emission, but what about absorption? The two weak features near velocities
of $-5$ and $-30$ \kms\ could at first sight be rejected as relatively
large noise wiggles. The uncomfortable fact, however, is that these
wiggles seem to occur close to velocities which have strong
corresponding \hto\ absorption, as can be seen in this
figure. The many absorption lines in the \hto\ profile
  originate in the different spiral arm features which the
  line of sight crosses between the Sun and the Galactic centre; there
is also broad \hto\ emission and self-absorption at positive velocities
originating in the +50 \kms\ Cloud itself. If the two weak features
in the upper (487 GHz \ot) profile were due to an unknown molecule with a
rest frequency near that of \ot, a different aligment with two other
\hto\ absorption lines would not agree for both lines due to the unique velocity
separation of this pair. A comparison of the
horizontal (H) and vertical (V) polarisation profiles may offer a little
support for the reality of these two weak features although the result
is just at the margin of detectability, as shown in Fig. 10. The
feature near
$-5$ \kms\ can be seen in both the H and V polarisations. The feature near
$-30$ \kms, although clear in the H profile, may be
masked by noise in the V profile and is certainly smaller than a
spurious wiggle at a more negative velocity, which, however, is
completely absent in the H profile. A careful study of
the OFF-spectra reveals a weak ripple-structure, which, however,
changes character at the positions of these two features. No obvious
emission could be identified in the OFF-spectra. Gaussian fits to the
two absorption features in Fig. 9 are presented in Table 2.  

In the following analysis, we shall hypothesize that these
two components are caused by the presence of \ot\ in the spiral
arms, absorbing the continuum radiation from the dust emission in the
+50 \kms\ Cloud at the Galactic centre. Since the {\it Herschel}
reference beams are offset from the main beam by 3\am\  along a PA of
92\degr, they observe regions outside the continnum emission from the +50
\kms\ Cloud (see Lis \& Carlstrom \cite{lis94}). The angular widths of
the spiral arms are much greater than this and we shall here
assume that the spiral arm gas is uniform in terms of the
excitation temperature and the
optical depth across the ON- and OFF-beams. The temperature in the
OFF-beams, $T_{\rm OFF}$,  will thus be given by 
\begin{equation}
T_{\mathrm{OFF}} = J(T_{\mathrm{ex}})(1-e^{-\tau}) + T_{\mathrm{sys}}
\end{equation}
where $J(T_{\rm ex}) = h\nu/k \times [{\rm e}^{h\nu/kT_{\rm
    ex}}-1]^{-1}$, $T_{\rm ex}$ is the excitation temperature, $\tau$ is the
optical depth, and  $T_{\rm  sys}$ is the system noise temperature. The
temperature in the ON-beam, $T_{\rm ON}$, is given by    
\begin{equation}
T_{\rm ON} = T_{\rm c}e^{-\tau} + J(T_{\rm ex})(1-e^{-\tau}) + T_{\rm
  sys}
\end{equation}
where $T_{\rm c}$ is the temperature of the continuum background. \\ 
The resulting double beam switched line intensity is then
\begin{equation}
\Delta T = T_{\rm ON} -  T_{\rm OFF} =  T_{\rm c}e^{-\tau}
\end{equation}
which allows us to determine the optical depth, $\tau$.

From our observations, we obtain a continuum antenna temperature of 119 mK at
487 GHz and using our absorption line values from Table 2, we
then get optical 
depths  of 0.029 and 0.021 for the $-7.1$ and $-32.6$ \kms\ features,
respectively. If we apply RADEX to the conditions in the cold neutral medium
with a temperature of 100 K and a density of $100$ cm$^{-3}$, the
previously determined optical depths result in column densites of
$1.3 \times 10^{18}$ cm$^{-2}$ in the $-7.1$ \kms\ feature and
 $1.5 \times 10^{18}$ cm$^{-2}$ in the $-32.6$ \kms\ feature.

No absorption lines are seen in the 774 GHz profile in
  Fig. 9. The continuum antenna temperature at this frequency is 288
  mK and the RMS noise level is 2.0 mK. Smoothing the profile to channel
resolutions of 4.7 and 7.5 \kms\ (which correspond to the line widths
of the $-7.1$ and $-32.6$ \kms\ 487 GHz \ot\ absorption features,
respectively) yield $3\sigma$ upper limits of $-2.7$ and $-2.1$ mK for
the $T_{\rm A}$ of possible absorptions. We thus have a lower limit of
about 1.2 for the ratio of intensities of the 487/774 GHz lines, which
implies that the kinetic temperature of the $-7.1$ and $-32.6$ \kms\
features must
be less than 60 K (RADEX, assuming a density of 1000 cm$^{-3}$, see
below). This temperature may be even lower, since we may here be
dealing with absorption lines rather than emission lines and the
energy of the lower level of the 774 GHz \ot\ line is 16.4 K. The
variation of the excitation temperature of the 487 GHz \ot\ line as a
function of density for three different kinetic temperatures is shown
in Fig. A.1 in the Appendix. (We note that excitation of the \ot\
molecule can occur by collisions with \HI\ and electrons as well as
with \htwo\  molecules (Gerin et al. \cite{ger15}; Lique et
al. \cite{liq14}).)

It is possible to limit the expected range of volume density of the
regions in question by using Sweden ESO Submillimeter Telescope (SEST)
observations of \ciso\ ($J=1-0$) 
(Lindqvist et al. \cite{lin95}) and \ciso\ ($J=2-1$) (Karlsson et
al. \cite{kar13}) at our position in the +50 \kms\ Cloud. We have
smoothed the ($2-1$) observations to the same angular resolution as the
($1-0$) observations, namely 45\as, which is similar to that of our
{\it Herschel} observations. The ratio of the two resulting profiles
for the $-7.1$ and $-32.1$ features is $\approx 1$. According to RADEX
modelling, this ratio requires molecular hydrogen densities of about 800 or
1800 cm$^{-3}$ for temperatures of 100 or 30 K, respectively. We
therefore conclude that the hydrogen density in these regions is closer
to 1000 cm$^{-3}$ rather than an order of magnitude lower.

 If we now again apply RADEX to the above absorption \ot\ optical
   depths using a temperature of 30 K and density of 1000 cm$^{-3}$ we
   get column densites of 
$7.9 \times 10^{17}$ cm$^{-2}$ and $9.0 \times 10^{17}$ cm$^{-2}$ for
the two features. The H$_2$  column densities of these two features
towards the +50 \kms\ Cloud are $5.5 \times 10^{21}$ and $3.2 \times
10^{21}$ cm$^{-2}$ (Sandqvist et al. \cite{san03}). Thus we arrive at
\ot\ abundances with respect to H$_2$ of
$X$(\ot) $= 1.4 \times 10^{-4}$ and $2.8 \times 10^{-4}$ in the $-7.1$
and $-32.6$ \kms\ features, respectively, implying that practically
all available oxygen is in the form of \ot. However, we note that
  the estimated strikingly high abundance of \ot\ partly is a result
  of the low \htwo\ column densities of the spiral arm clouds
  (corresponding to only 3-6 magnitudes of visual extinction). We
  estimate an uncertainty in the H$_2$ values of $40 \%$, of which
  $30 \%$ is due to the uncertainty in the CO-H$_2$ conversion factor
  in the Milky Way disk (Bolatto et al. \cite{bol13}).

The corresponding main beam brightness temperature 
of the 487 GHz \ot\ emission from the uniform, extended spiral arm cloud
assumed in the current analysis, as predicted by RADEX, is $0.2$
K. However, the 5$\sigma$\ intensity limit reached in our earlier {\it Odin} 
observations at 487 GHz, using position-switching to a distant,
emission-free region is as low as 0.02 K in the $2\farcm4$ antenna beam
(Sandqvist et al. \cite{san08}). This leads us to conclude that our previous
assumption of an exactly uniform cloud covering the {\it Herschel} ON and
OFF beams must be incorrect. It is unlikely that the \ot\
  absorption comes from the cold neutral medium, but rather from more
  compact molecular clouds. 

One way to avoid the above contradiction (perhaps unlikely) would be
to assume that the {\it Herschel} lines are caused by localized
absorption in front of the continuum source, i.e. the regions giving
rise to these lines are small with respect to the {\it Odin} beam. This
would result in an {\it Odin} non-detection due to beam dilution
effects and the absorbing clouds would have to be smaller than $68\as
- 46\as$.   

\subsection{\ot\ upper limits implied by {\it Odin} observations}

Based on the assumption that there are no beam dilution effects, the two weak
  \ot\ absorption lines observed by {\it Herschel}, if at all real,
  may instead be caused by a weak excess emission in the OFF beam
average, as compared to that in the ON position, although we have not 
been able to confirm this from our study of the OFF-spectra.  
If we apply RADEX, with the same cloud parameters as above (30 K and
1000 cm$^{-3}$), to the 5$\sigma$\ upper limit of 0.02 K observed by
{\it Odin} in the ON position, the \ot\ column density upper limits in
the $-7.1$ and $-32.6$ \kms\ features instead become $\le 4.9 \times
10^{16}$ cm$^{-2}$ and $\le 7.8 \times 10^{16}$ cm$^{-2}$,
respectively. Thus we arrive at upper limits to the \ot\ abundances in
these two features of $\le 9 \times 10^{-6}$ and $\le 2 \times 10^{-5}$.

\subsection{{\it Herschel} differential OFF-ON \ot\ emission}

If we now assume that the observed {\it Herschel}
  absorption lines are apparent only and due to a differential
  emission between the OFF and ON beams we may obtain indications of
  \ot\ abundance limits depending upon the specific conditions. We
  consider it unlikely that there will be emission in both OFF beams
  and none in the ON beam, so the simplest condition is no
  emission in the ON beam and emission in one OFF beam. In this case
  we apply RADEX to the same conditions as before, i.e. for a
  temperature of 30 K and a density of 1000 cm$^{-3}$.  Correcting for
  a beam efficiency of 0.62, we get main beam brightness temperatures
  of +5.5 and +4.0 mK for the $-7.1$ and $-32.6$ \kms\ ``emission''
  features, respectively. But
  since we assume emission in only {\it one}\ OFF beam, the effective
  main beam brightness temperatures will be twice as high, i.e. +11.0
  and +8.0 K. The best RADEX fits then give \ot\ column
  densities of $2.7 \times 10^{16}$ and $3.1 \times 10^{16}$
  cm$^{-2}$, which correspond to \ot\ fractional abundances of
  $4.9 \times 10^{-6}$ and $10 \times 10^{-6}$, respectively for the two
  features. These two \ot\ abundances can be considered as {\it
    Herschel} upper limits and are less than the previous {\it Odin}
  values, see Table 3. A more complicated condition would be if there
  {\it is} emission also in the ON beam, 
  then the OFF emission must be stronger to compensate for the ON
  emission, and thus the above limits would really be lower limits.

A summary of the \ot\ abundance values is presented in Table 3, where
we have also tabulated the corresponding abundances of OH, \hto, and
\nht\ resulting from Very Large Array (VLA), SEST, and {\it Odin}
observations (Karlsson et al. \cite{kar13}).  

\subsection{\ot\ and Galactic spiral arm shocks}

The highest  \ot\ abundances modelled so far to fit
observations were developed to explain the observations of \ot\  in a
unique highly compact region in Orion, where a value of close to
$10^{-4}$ could be obtained by means of shock chemistry (Chen et al. 2014). 
The velocity of the feature near 0 \kms\  is $-7.1$ \kms, which probably
means that it originates in the giant Scutum spiral arm (Sandqvist
1970). The \htco\ profile in that paper, which is also reproduced in
Fig. 8, has an exceptionally deep absorption close to this
velocity. The other feature has a velocity of $-32.6$ \kms\ and thus
probably originates in the $-30$ \kms\ Arm. The Scutum Arm is one of
the two {\it major} spiral arms in the Galaxy and in the direction of
the Galactic centre lies at a distance of $\approx$3 kpc from the Sun
(Reid et al. 2009). The $-30$ \kms\ Arm is even closer to the central
region of the Galaxy, but still outside the 3 kpc Arm (e.g. Corbel \&
Eikenberry 2004). We note in Table 2 that the velocities of the
  \chl\ features differ from the \ot. The \chl\ feature near $-51$ \kms\
  originates in the 3 kpc Arm, that near $-3$ \kms\ in the Sgr Arm and
  that near +4 \kms\ in the Local Arm. Also, the U1(HDCO?) line is
  detected only in the Local and Sgr arms, possibly due to blending
  effects.  On the other hand, \htco\ is apparently present in all of
  the Local, Sgr, Scutum, $-30$ \kms, and 3 kpc Arms (Fig. 8 and
  Sandqvist \cite{san70}).

What makes the Scutum Arm and the $-30$ \kms\ Arm possibly conducive to a
high abundance of \ot\ is at the present time a mystery if the
two absorption features are indeed due to \ot. However, none of
  the other spiral features along the line of sight to the Galactic
  centre is even comparable in magnitude to the Scutum Arm.
A study of the shock chemistry caused by the action of the Galactic
density wave (e.g. Shu et al. \cite{shu73}; Roberts \& Stewart
\cite{rob87}; Sundelius et al. \cite{sun87}; Martos \& Cox 1998; Gomez
\& Cox 2004a,b) in the major Scutum spiral arm  
might provide some answers to our dilemma. The existence of spiral
density wave shocks in the Milky Way galaxy is very likely but has not
been fully observationally proven because of our location inside the
Galaxy. However, the existence can safely be inferred, based upon the
convincing observational proof in some nearby spiral galaxies. In the
grand design Sc spiral galaxy M51, both the radial and tangential
velocity components, as observed in CO, show steep gradients
across the spiral arms in accordance with the predictions of density
wave theory and numerical modelling experiments (e.g. Rydbeck et
al. \cite{ryd85}; Adler et al. \cite{adl92}; Aalto et al. \cite{aal99};
Schinnerer et al. \cite{sch10}). Similar convincing results have
been published in case of the SAB galaxy M83 (Lundgren et al
2004a,b). In addition, strong shocks have been predicted and observed
along the leading edges of galactic bars, e.g. NGC 1365 (Lindblad et
al. \cite{lin96}). This galaxy presents strong absorbing lanes and
strong velocity shocks along the leading side of the bar, but also a
number of dark lanes across the bar ending at the shocks. 

\begin{table*}
\caption{Column densities and abundances for \ot. ($T_{\rm k}=30$ K,
  $n_{\rm H} = 1000$ cm$^{-3}$)}
\begin{flushleft}
\begin{tabular}{llllllll}
\hline\hline\noalign{\smallskip}
Feature & Origin & $N$(\ot) & $N$(\htwo) & $X$(\ot) &
$X$(OH) \tablefootmark{a} & $X$(\hto) \tablefootmark{a} &
$X$(\nht) \tablefootmark{a}  \\ 
(\kms]    &   &   (cm$^{-2}$) &  (cm$^{-2}$)  & $(\times 10^{-6})$  &
$(\times 10^{-6})$ & $(\times 10^{-9})$ & $(\times 10^{-9})$ \\  
\hline\noalign{\smallskip}
+45.2  &  +50 \kms\ Cloud &  $\le 1.4 \times 10^{16}$ & $2.4
\times 10^{23}$ & $\le 0.05$  & $8$ & $40$ &  \\
$-7.1$ & Scutum Arm &  $7.9 \times 10^{17}$ & $5.5 \times
10^{21}$ & $140$ \   \tablefootmark{b}  & $2$ \   \tablefootmark{e}
&   & $3$ \   \tablefootmark{e} \\ 
   &  & $\le 4.9 \times 10^{16}$ & & $\le 9$  \ \tablefootmark{c} &   &   & \\
   &  & $2.7 \times 10^{16}$ & & $4.9$ \ \tablefootmark{d} &
   &   & \\ 
$-32.6$ & $-30$ \kms\ Arm & $9.0 \times 10^{17}$ & $3.2 \times
10^{21}$ & $280$ \   \tablefootmark{b} & $6$  & $30$  & $5$ \\  
   &  & $\le 7.8 \times 10^{16}$ & & $\le 20$  \ \tablefootmark{c} &   &   & \\ 
   &  & $3.1 \times 10^{16}$ & & $10$ \ \tablefootmark{d} &
   &   & \\    
\noalign{\smallskip}\hline\end{tabular}
\end{flushleft}
$^a$ from Karlsson et al. (\cite{kar13}) \\
$^b$ {\it Herschel} ``ON-absorption''\\
$^c$ {\it Odin} limit \\
$^d$ {\it Herschel} ``differential OFF-ON emission'' \\
$^e$ may be a blend of the Local, Sagittarius, and Scutum Arms \\
\end{table*}

If we assume that shocks are responsible for the \ot\ tentatively
  detected in absorption at $-7.1$ \kms\ and at $-32.6$ \kms, then using the
  models presented in Chen et al. (\cite{che14}) we find that we are only able
  to produce enough \ot\ by assuming a pre-shock density of 1000 cm$^{-3}$
  and a shock velocity of at least 20 \kms. These may not be
    unreasonable values for the Milky Way galaxy. In M51, a galaxy
    with a strong spiral density wave, the value of 
    arm-to-interarm CO intensity ratio of $\approx 6$, found by Aalto
    et al. (\cite{aal99}), implies the existence of some moderately
    high density interarm
    gas which enters the density wave shock  and spiral arm region with a
    velocity of $\approx 50$ \kms. Of course, as reported in 
  Chen et al. (\cite{che14}), several assumptions are made when running a 1D
  shock chemical model, including the geometry of the shocked
  regions. However, our tentative detection cannot be substantiated by
  any further modelling at this stage and until further observations
  confirm the presence of \ot. 

\subsection{Possible alternative interpretation of the absorption
  features} 

It should also be mentioned that a frequency coincidence of these two
absorption features (if near 0 and $-30$ \kms) exists with the cis-HOCO
($4_{2,2} - 3_{1,3}$) line at 487.256 GHz. The energy of the lower state is
only 12.7 K. cis-HOCO is a higher energy conformer (800 K
higher than trans-HOCO) of a not yet detected interstellar
molecule. However, even a tentative identification of cis-HOCO is
ruled out by the fact that the line in question is only one out of ten
cis-HOCO ($4_{2,2} - 3_{1,3}$) hyperfine transitions, where the
transitions at 487.535, 487.536, 488.195 and 488.201 GHz are expected
to be more than an order of magnitude stronger than the one at 487.256
GHz. These other hyperfine transitions are not visible in our observed
spectrum (the Einstein $A$ coefficients of the four lines are 30 times
larger than that of the identification candidate).

\begin{figure}
  \resizebox{\hsize}{!}{\rotatebox{270}{\includegraphics{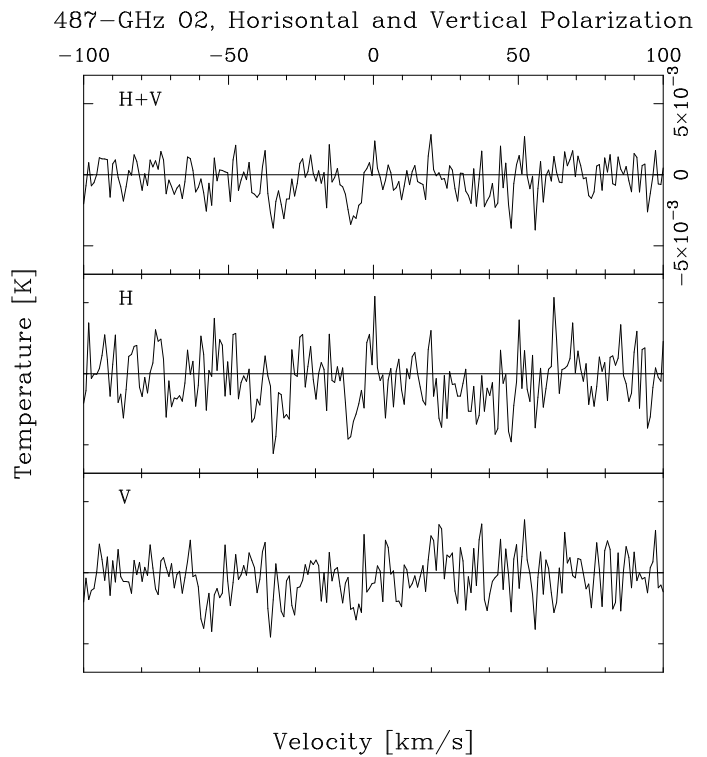}}}
  \caption{487 GHz \ot\ spectra towards the \sgra\ +50 \kms\
    Cloud. Horizontal ({\it middle} H) and Vertical ({\it lower} V)
    polarisations. The average ({\it upper} H+V) of these two profiles
    is also shown.}     
  \label{10}
\end{figure} 

\subsection{Methanol properties of the +50 \kms\ Cloud}

We have made a simple population diagram analysis (Goldsmith \&\
Langer 1999) of unblended \met-$A$ and \met-$E$ lines in Table 1 (the \metiso\
lines gave inconclusive results, possibly due to sensitivity
effects and blends). The diagram is shown in Fig. 11. The slopes of the fitted
straight lines gave rotation temperatures of $T_{\rm rot} \approx 64$
and 79 K for \met-$A$ and \met-$E$, respectively, at the
observed position of the +50 \kms\ Cloud. This range is close to the
kinetic temperature of 80 K which we have adopted for this cloud. A
further RADEX analysis of the \met-$A$ lines, using the same
input values as for the above \ot\ RADEX analysis, yielded a best fit
column density of $N$(\met) $ \approx 1.3 \times 10^{17}$ cm$^{-2}$, and thus an
abundance with respect to H$_2$ of $X$(\met) $ \approx 5 \times
10^{-7}$. This value is slightly greater than the upper value in the
range of $10^{-7} - 10^{-9}$ found generally in the  inner 30 pc of
the Galactic centre (e.g. Yusef-Zadeh et al. 2013). But it is
consistent with the higher production rate of \met\ expected from the
proximity and interaction of the +50 \kms\ Cloud with the high-energy
expanding non-thermal shell of \sgraeast.  

\begin{figure}
  \resizebox{\hsize}{!}{\rotatebox{270}{\includegraphics{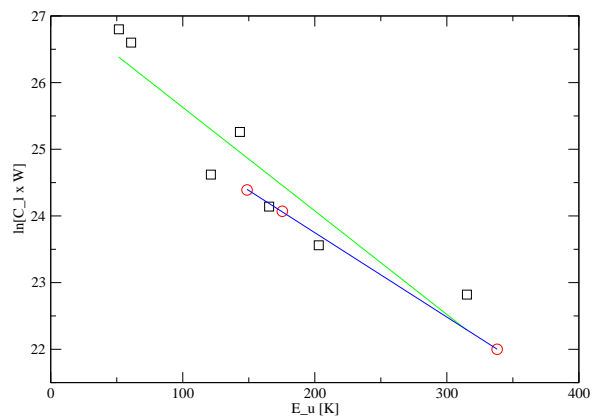}}}
  \caption{Population  diagram for \met-$A$ (squares) and \met-$E$
    (circles). $E_u$ is the energy of the upper level, $C_l$ a
    constant of the line and $W$ is the integrated main beam line
    intensity.}      
  \label{11}
\end{figure}

\section{Conclusions}

We have searched for \ot\ in the \sgra\ +50 \kms\ molecular
cloud in the Galactic centre, using the HIFI system aboard the {\it
  Herschel} Space Observatory. We obtain a $3\sigma$ upper limit
for the \ot\ abundance relative to \htwo\ of $X$(\ot) $\le 5 \times
10^{-8}$ in that cloud. From an analysis of the \met-$A$ and \met-$E$
emission lines we obtain a rotation temperature range of approximately
64 to 79 K and a \met\ abundance of $5 \times 10^{-7}$. 

If two {\it Herschel} weak apparent absorption lines in the 487 GHz
spectrum are due to \ot, we estimate the \ot\
abundance with respect to H$_2$ in two inner Galactic spiral arms (the
Scutum Arm and the $-30$ \kms\ Arm) to be 

(i) $1.4 - 2.8 \times 10^{-4}$, assuming absorption by foreground clouds,

(ii) $\le (1 - 2) \times 10^{-5}$, assuming weak excess emission in
the {\it Herschel} OFF-beam average and using {\it Odin}
non-detection limits, 

(iii) $\approx (5 - 10) \times 10^{-6}$, assuming that the apparent
absorption lines are due to a differential emission between the
{\it Herschel} OFF and ON beams and using {\it Herschel} intensities.

A simple model study suggests that shocks caused by a 
Galactic spiral density wave may produce \ot\ in a 30 K and 1000
cm$^{-3}$ medium located in the giant Scutum spiral Arm.

\begin{acknowledgements} 

We wish to thank Per Olof Lindblad for useful discussions and comments
concerning Galactic density waves and resulting shock interaction with
Galactic spiral arms. We are very grateful to our anonymous referee
  whose thorough report significantly improved this paper. We also wish to
thank the Swedish National Space Board (SNSB) for its continued financial
support. Furthermore we express our appreciation to the individuals
making, updating and maintaining the Splatalogue, CDMS and JPL
molecular spectroscopy data bases for their unselfish demanding
work. HIFI has been designed and built 
by a consortium of institutes and university departments from across Europe,
Canada and the US under the leadership of SRON Netherlands Institute for Space
Research, Groningen, The Netherlands with major contributions from Germany,
France and the US. Consortium members are: Canada: CSA, U. Waterloo;
France: CESR, LAB, LERMA, IRAM; Germany: KOSMA, MPIfR, MPS; Ireland,
NUI Maynooth; It aly: ASI, IFSI-INAF, Arcetri-INAF; Netherlands: SRON,
TUD; Poland: CAMK, CBK; Spain: Observatorio Astronomico Nacional
(IGN), Centro de Astrobiologia (CSIC-INTA); Sweden: Chalmers
University of Technology MC2, RSS \& GARD, Onsala Space Observatory,
Swedish National Space Board, Stockholm University Stockholm
Observatory; Switzerland: ETH Zürich, FHNW; USA: Caltech, JPL,
NHSC. Support for this work was provided by NASA through an award
issued by JPL/Caltech.

\end{acknowledgements}

\appendix
\section{Excitation temperature of the 487 GHz \ot\ line}

The detection of \ot\ in absorption would be in itself surprising. To
see an absorption line, the excitation temperature must be lower
than the background contiuum temperature. But the \ot\ lines have low
critical densities and are easily excited even in low density
foreground clouds. We have performed a RADEX analysis to
illustrate the variation of excitation temperature of the 487 GHz \ot\ line as a
function of molecular hydrogen cloud density at three kinetic
temperatures, viz. 30, 60 and 100 K. The results are shown in Fig. A.1.

\begin{figure}
  \resizebox{\hsize}{!}{\rotatebox{90}{\includegraphics{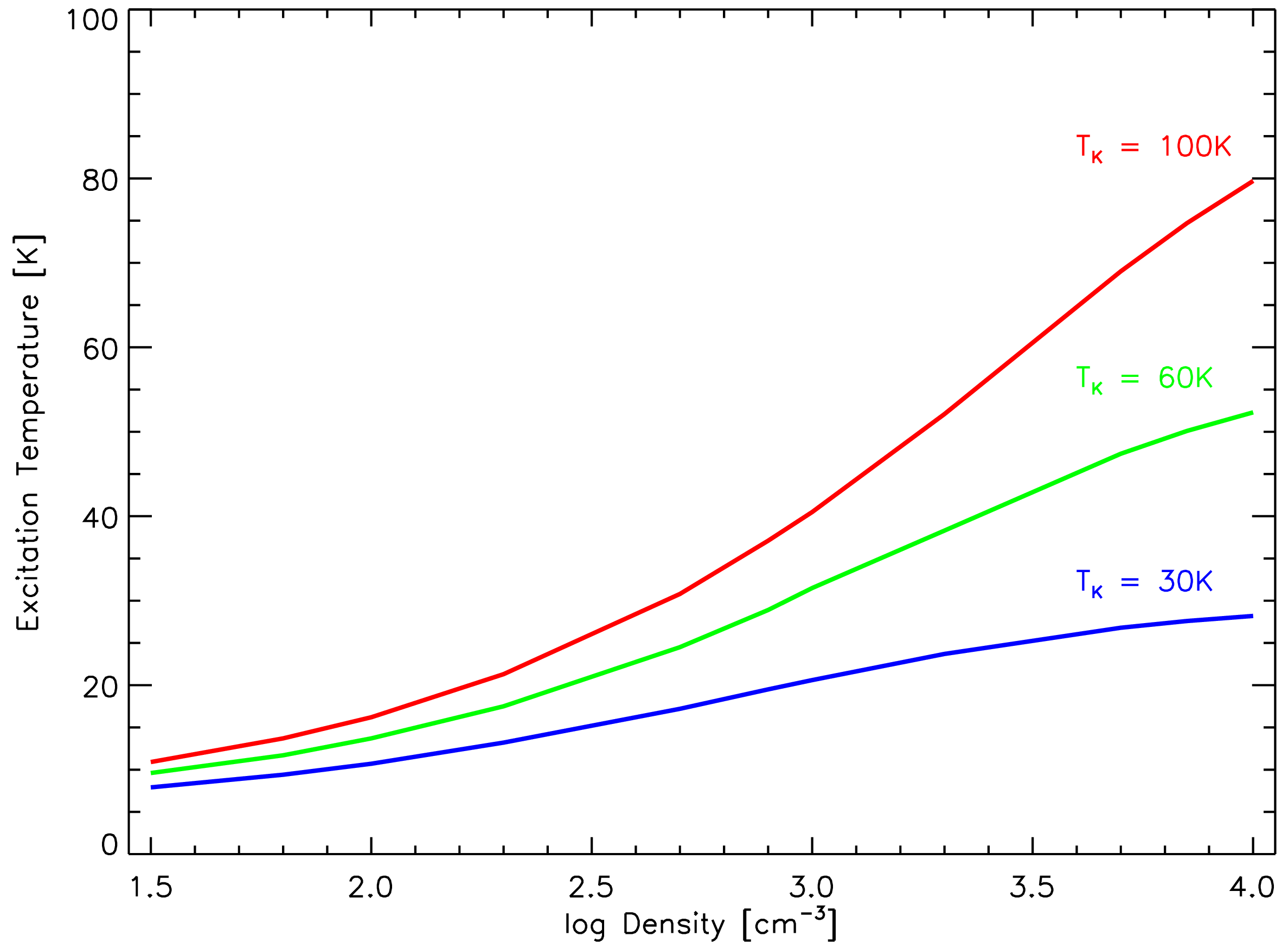}}}
  \caption{Excitation temperature vs. cloud density for the 487 GHz
    \ot\ transition resulting from RADEX multi-transition analysis,
    using three values of kinetic temperature: 30, 60 and 100 K.}    
  \label{A.1}
\end{figure}

\end{document}